\documentclass[aps,pra,reprint,showpacs,superscriptaddress]{revtex4-2}
\usepackage{bm,amsfonts,amssymb,amsmath,mathrsfs}
\usepackage{graphicx,subfigure}
\usepackage{float}
\usepackage{xcolor}
\usepackage[colorlinks,linkcolor=blue,citecolor=blue,urlcolor=blue]{hyperref}
\usepackage{physics}

\usepackage{bbold}
\usepackage{color}

\usepackage{amsfonts,amsmath}
\usepackage{epsfig,amsmath}
\usepackage{graphicx}
\usepackage{dcolumn}
\usepackage{bm}
\usepackage{bbold}
\newcommand{\beq}{\begin{equation}}
\newcommand{\eeq}{\end{equation}}
\newcommand{\beqa}{\begin{eqnarray}}
\newcommand{\eeqa}{\end{eqnarray}}

\newcommand{\la}{\langle} 
\newcommand{\ra}{\rangle}

\makeatletter
\newcommand*{\revision}{\@ifnextchar\bgroup{\revision@}{\color{red}}}
\newcommand*{\revision@}[1]{{\textcolor{black}{#1}}}
\makeatother

\begin{document}

\title{Wave-particle duality ellipse and application in quantum imaging with undetected photons}

\author{Pawan Khatiwada}
\author{Xiao-Feng Qian}
\email{xfqian6@stevens.edu}
\affiliation{Center for Quantum Science and Engineering, and Department of Physics, Stevens Institute of Technology, Hoboken, New Jersey 07030, USA} 

\begin{abstract}
{We present a systematic framework to quantify the interplay between coherence and wave-particle duality in generic two-path interference systems. {Our analysis reveals a closed-form duality ellipse (DE) equality, that rigorously unifies visibility (a traditional waveness measure) and predictability (a particleness measure) with degree of coherence, providing a complete mathematical embodiment of Bohr’s complementarity principle.} Extending this framework to quantum imaging with undetected photons (QIUP), where both path information and photon interference are inherently linked to spatial object reconstruction, we establish an imaging duality ellipse (IDE) that directly connects wave-particle duality to the object’s transmittance profile. This relation enables object characterization through duality measurements alone and remains robust against experimental imperfections such as decoherence and misalignment. Our results advance the fundamental understanding of quantum duality while offering a practical toolkit for optimizing coherence-driven quantum technologies, from imaging to sensing.}
\end{abstract}

\maketitle

\section{Introduction}

{Wave-particle duality sits at the heart of quantum mechanics, directly inspiring Schr\"odinger's development of of his wave equation to account for de Broglie's ad hoc matter wave theory from first principle \cite{de1924recherches,bohr1928quantum,bloch1976heisenberg}.} The quantitative characterization of waveness and particleness by Wootters and Zurek \cite{wootters1979complementarity} led to the formulation of different versions of a remarkable duality inequality \cite{Glauber1986,greenberger1988simultaneous,jaeger1993complementarity,jaeger1995two, englert1996fringe,liu2012relation,qian2020quantumduality,yoon2021quantitative}, $V^2+D^2\le 1$, where visibility $V$ and distinguishability (or predictability) $D$ quantify respectively the waveness and particleness of a quantum object. However, this inequality does not always reflect the mutually exclusive relationship between wave and particle characteristics, as an increase in one does not necessarily lead to a decrease in the other, which contradicts Bohr's complementarity principle \cite{bohr1928quantum}. {The reduction from a fully complementary equality to an inequality is primarily due to partial coherence of the two path states \cite{greenberger1988simultaneous,jaeger1993complementarity}. Such partial coherence will inevitably induce entanglement between the path and remaining (external) degrees of freedom. Therefore, important advancements have been made to include the role of entanglement into the wave-particle duality picture \cite{jakob2010quantitative,qian2018entanglement,norrman2017complementarity,de2018hidden,basso2020complete,qureshi2021predictability}. Particularly, coherence has been treated as an enabler for entanglement in connection with predictability/distinguishability and wave interference visibility \cite{bera2015duality,basso2022entanglement}. The extended role of coherence and entanglement have also been explored in uncertainty relation \cite{durr2001quantitative}, Lorentz invariance \cite{basso2021complete}, entanglement swapping \cite{basso2022entanglementpla}, polarization coherence \cite{qian2023bridging,Yang2025photonic}, etc. However, in most of these studies, coherence is embedded in other quantities, and the direct role of partial coherence and its precise impact on the duality relation has not been fully explored. This raises the question: what is the fundamental quantitative role of coherence in quantum duality?}

{Furthermore, wave and particle features, along with their constraints, are recognized as crucial elements in the execution of various quantum measurement, sensing, imaging, control tasks \cite{white1998interaction,lemos2014quantum,qian2020turning,das2020wave,wang2022controlling,janovitch2023wave,li2023experimental}.} However, a comprehensive quantitative analysis of the interplay between these dual aspects (as captured by duality relations) in specific quantum tasks remains largely unexplored. {This leads to a key follow-up question: what is the operational role of wave-particle duality, quantitatively mediated through coherence, in quantum tasks?}

To address these questions, we quantitatively analyze how generic two-path coherence constrains wave-particle duality in the Mach-Zehnder interferometer scenario. By incorporating the degree of coherence (via standard optical cross-correlation \cite{bornwolf,mandel1991coherence}), a compact duality ellipse (DE) relation is obtained. Coherence is shown to be dictating the DE’s ellipticity, governing the wave-particle tradeoff. {The Duality Ellipse resolves ambiguities in the duality inequality, ensuring strict adherence to Bohr’s complementarity principle by quantitatively enforcing the mutual exclusivity of wave-like (waveness) and particle-like (particleness) properties through the integration of coherence.}

We then extend our analysis to account for generic forms of coherence loss and explore the application of the derived DE relation in the recently emerging field of quantum imaging with undetected photons (QIUP) \cite{lemos2014quantum,lahiri2015theory,gilaberte2021video,topfer2022quantum,santos2022subdiffraction,haase2023phase,Dalvit2024,machado2024complementarity}. In the QIUP scenario, both wave and particle characteristics of the photon are crucial to achieving imaging objectives. Previous studies have highlighted the importance of indistinguishability \cite{lahiri2015theory,lemos2022quantum,hochrainer2022quantum} as a manifestation of waveness in the imaging process. However, the collective influence of particle nature, particularly the role of the quantitative duality relation, remains largely unexplored. Our approach involves a quantitative integration of object information, represented by its transmittance $T$ (a generalized form of coherence), into the dual facets of waveness (quantified by interference visibility $V$) and particleness (measured by predictability $D$ \cite{greenberger1988simultaneous,jaeger1993complementarity}). We demonstrate that when a photon is absorbed by the object, it introduces particle information, leading to the inevitable loss of coherence and a subsequent reduction in interference visibility. This particle information, crucial for object probing, is quantitatively incorporated to construct the image of an object. The incorporation of object information results in an extended imaging duality ellipse (IDE) relation, where the ellipticity (in terms of $T$) can be used to obtain object information. In essence, QIUP is demonstrated to be quantitatively governed by the photon's fundamental wave-particle duality, suggesting an alternative approach for obtaining object information through measuring the waveness and particleness of the photon.

\section{Quantum Duality Ellipse} 
\begin{figure}[h!]
\begin{center}
    \includegraphics[scale=1]{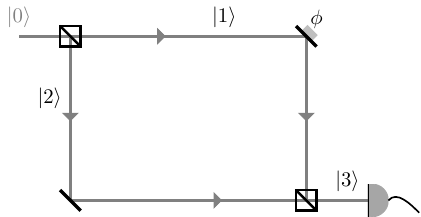}
    \caption{Schematic of the Mach-Zehnder interference setup for a single photon to analyze wave and particle features.}
    \label{Fig:SimpleSetup}
\end{center}
\end{figure}

We consider the standard two-path scenario by analyzing the typical example of a photon passing through a Mach-Zehnder interferometer (MZI) as illustrated in Fig.~\ref{Fig:SimpleSetup}. While the main results and conclusions hold for any quantum particle, we conduct the following analysis using a photon for convenience, without any loss of generality. A photon starts with the state $|0\ra|M_0\ra$, where $|0\ra$ represents the initial path mode and $|M_0\ra$ describes the combined state of the remaining internal degrees of freedom of the photon along with all external states of the rest of the universe. Then it enters the MZI and splits into two path states $|1\ra$ and $|2\ra$ with a generic probabilities of $P_1$ and $P_2$ respectively. When recombined by the second BS, the two path states converge into the same $|3\ra$ path mode, where ideal interference happens if the marginal state $|M_0\ra$ remains unchanged during the process. 

In realistic scenarios, decoherence or other mechanisms such as photon loss can modify the reduced state of the photon’s path modes, degrading coherence. Then the (environmental) marginal states in path 1 and path 2 evolve as $|M_0\ra \rightarrow |M_1\ra$ in path 1 and $|M_0\ra \rightarrow |M_2\ra$ in path 2. Therefore, the most general form
of the photon path mode state after the first BS can always be written as 
\beq
|\Psi_1\ra=c_1 |1\ra |M_1\ra + c_2 |2\ra|M_2\ra,
\label{initial state}
\eeq
where $|c_1|^2=P_1$ and $|c_2|^2=P_2$ are the normalized coefficients with $P_1+P_2=1$. Then the coherence of the two paths 1 and 2 is determined by the normalized states $|M_1\ra, |M_2\ra$. Here we adopt the optical definition of degree of coherence $\gamma$ in terms of the normalized cross-correlation of $|M_1\ra$ and $|M_2\ra$ \cite{bornwolf}, i.e.,
\beq 
\gamma=|\la M_1|M_2\ra|,
\eeq
which varies between 0 and 1 with 0 indicating complete incoherence and 1 meaning fully coherent. {Note that this degree of coherence is also quantitative related to entanglement (measured by concurrence $C$ \cite{wootters1998entanglement}) between the path and marginal degrees of freedom, i.e., $C=2|c_1c_2|\sqrt{1-\gamma^2}$. Therefore, our following analysis of coherence is naturally also connected to the quantitative links between duality and entanglement \cite{qian2018entanglement,de2018hidden, basso2020complete,qureshi2021predictability,basso2022entanglement}.}

After the second BS (50/50), the two paths converge into the same path mode $|3\ra$. Then the photon (non-normalized) state becomes
\begin{equation}
    \label{interference}
    \ket{\Psi_2} \propto  c_1\ket{3}|M_1\ra+ e^{-i\phi}c_2\ket{3}|M_2\ra,
\end{equation}
where $\phi$ is the relative phase accumulated between the two paths. The probability $P$ of the photon at the detector is achieved straightforwardly as $P \propto \abs{c_1}^2 + \abs{c_2}^2 + 2\abs{c_1c_2}\gamma \cos(\phi)$. {Then the interference visibility $V$ and predictability $D$ can be computed respectively \cite{greenberger1988simultaneous, jaeger1995two} as $V = (P_{max}-P_{min})/(P_{max}+P_{min})=2\abs{c_1c_2}\gamma$ and $D=\abs{P_1-P_2}= \abs{\abs{c_1}^2-\abs{c_2}^2}$.} From the schematic setup in Fig.~\ref{Fig:SimpleSetup}, $V$ can be achieved by registering photon counts when continuously varying the phase $\phi$ through the piezoelectric transducer in path $\ket{1}$, and $D$ can be obtained by measuring $P_1$, $P_2$ when blocking paths $|2\ra$, $|1\ra$ respectively. {While these two quantities apparently lead to the conventional duality inequality \cite{greenberger1988simultaneous, jaeger1995two}, i.e., $V^2+D^2\le 1$, it also leads to an exact closed relation
\beq
\label{Duality Ellipse}
 \frac{V^2}{\gamma^2}+D^2=1,
\eeq
where the quantitatively role of coherence is now apparent in terms of $\gamma$.}

\begin{figure}[h!]
    \centering
    \includegraphics[scale=1]{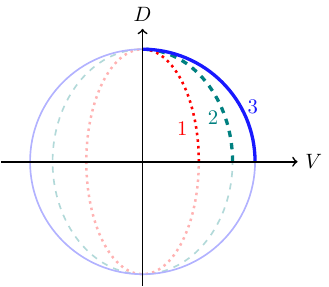}
    \caption{Illustration of the duality ellipse for ellipticity $\eta = 1-\gamma=1-T =  0$, (blue, curve 3), 0.2 (teal, curve 2), and 0.5 (red, curve 1) respectively. Curve 3 represents the case when the two paths are completely coherent, i.e., $\gamma=1$ or no object in the QIUP with $T=1$.}
    \label{Circlellipse}
\end{figure}

{When $V$ and $D$ are treated as two independent coordinates, this duality relation describes an exact ellipse equation, see illustration in Fig.~\ref{Circlellipse}. The degree of coherence $\gamma$ is the reason causing the elliptical nature of the duality relation as the ellipticity is exactly the complement of coherence $\eta = 1-\gamma$. Thus, we call the above result \eqref{Duality Ellipse} the Duality Ellipse (DE). }

{Apparently, through the relation $C=2|c_1c_2|\sqrt{1-\gamma^2}$, one can directly connect our relation to the alternative sets of complete complementarity relations that focus on the role of entanglement \cite{qian2018entanglement,de2018hidden, basso2020complete,qureshi2021predictability,basso2022entanglement}. However, our Duality Ellipse relation explicitly singles out the role of coherence $\gamma$. More deeply, our relation suggests that the adjusted interference visibility $V$ by coherence $\gamma$, i.e., $V/\gamma$, serves as a perfect new measure of waveness to ensure a rigorous complementary equality. It avoids the ambiguity of the exclusiveness between waveness and particleness, thus exactly compatible with Bohr's complementarity statement \cite{bohr1928quantum}. }

{A critical feature of our Duality Ellipse is its ability to quantitatively unify coherence effects from both intrinsic (internal) and environmental (external) degrees of freedom, all included in the marginal states $\gamma=|\la M_1|M_2\ra|$. This universality ensures the DE applies equally to systems in pure and mixed states. Notably, in the fully coherent limit $\gamma=1$, the DE reduces to the conventional pure-state equality $V^2+D^2=1$, thereby recovering the well-established duality relation as a special case of our formalism.}

It is worth to note that the partial coherence parameter $\gamma$ also accounts for photon loss scenarios. For example, if a photon in path 1 is subject to loss (e.g., via interaction with environment), the joint system-environment state becomes $|1\rangle|M_1\rangle \rightarrow a|1\rangle|M_1\rangle+ b|l\rangle|M_l\rangle$, where $|l\ra$ denotes the lost-photon state with probability $|b|^2$. This modifies the coefficient $c_1$ in Eq.~\eqref{initial state} to becomes $c'_1=c_1 a$. Crucially, our duality analysis remains fully valid under this substitution, as the derivation depends only on the structure of the coefficients and not their specific origin.


\section{Imaging Duality Ellipse}
\begin{figure}[h!]
    \centering
    \includegraphics[scale=1]{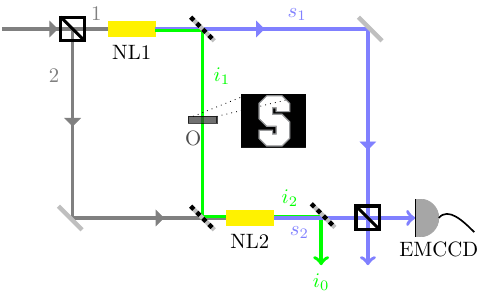}
    \caption{ Schematic illustrations of the QIUP setup in Ref.~\cite{lemos2014quantum}. Here, object $O$ is represented by the cut-out letter ``S''.}
    \label{Setup1}
\end{figure}

We now consider the application of the obtained DE (\ref{Duality Ellipse}) to the scenario of quantum imaging with undetected photon (QIUP) \cite{lemos2014quantum}, which is based on the Zou-Wang-Mandel (ZWM) scheme of induced coherence without induced emission \cite{Zou1991PRL}, see a schematic illustration in Fig.~\ref{Setup1}.

As is shown in Fig.~\ref{Setup1}, an input pump beam is split by a general beamsplitter (represented by the cube with the slanted lines) into two, both of which are then down converted by a corresponding identical Non-Linear (NL) crystal (illustrated by a yellow box) into a pair of respective entangled photons \cite{ou2017quantum}. The green and blue lines represent the idler and signal photon paths respectively. The first idler photon mode, $i_1$ (generated from NL1), is then reflected or absorbed partially by object $O$. The transmitted portion is then aligned with the idler photon mode $i_2$ of the second crystal NL2 to become a common idler photon mode $i_0$ that exits the interferometer. The signal photon modes, $s_1,s_2$ from the two corresponding crystals are then combined by a 50/50 beamsplitter (BS) for interference detection. 

We first analyze the sequence of quantum states along the optical path in this QIUP setup by temporarily assuming that, (a) there is no mechanism of coherence loss except for the existence of the object, and (b) there is perfect alignment of the two idler modes. We will release the two restrictions in the following section. 

Under assumption (a), the pump photon path states are completely coherent after split into two by the first BS, and can be represented as $\ket{\psi_0} = c_1\ket{1} + e^{-i\phi_0}c_2 \ket{2}$, where $c_1$ and $c_2$ are normalized coefficients, with $\abs{c_1}^2+\abs{c_2}^2=1$. The two path states are denoted as $\ket{1}\equiv|1\ra_{p_1}|0\ra_{p_2}$ and $\ket{2}\equiv|0\ra_{p_1}|1\ra_{p_2}$ with $|1\ra_{p_1}$ and $|0\ra_{p_1}$ indicating respectively 1 and 0 photon in path $p_1$, etc. The relative phase $\phi_0$ is induced due to reflection and path length difference. 

After spontaneous parametric down conversion (SPDC) \cite{mandel1995optical} in NL1 and NL2, the photon state becomes
\begin{equation}
    \label{SPDC-state}
    \ket{\psi_1} =  c_1\ket{s_1}\ket{i_1}+  e^{-i\phi_1} c_2\ket{s_2}\ket{i_2} ,
\end{equation}
where $\phi_1$ is the new relative phase after the non-linear crystals. Here $\ket{s_{1,2}}$ and $\ket{i_{1,2}}$ represent the signal and idler photons down converted from the non-linear crystals NL1,2 respectively. The single-photon pair state is selected with other photon-number states discarded. Note that a more complete description of the state should be $\ket{s_1}\ket{i_1}\equiv\ket{0}_{p_1}\ket{1}_{s_1}\ket{1}_{i_1}\ket{0}_{p_2}\ket{0}_{s_1}\ket{0}_{i_1}$ where $\ket{1}_{s_1}$ denotes 1 photon in the signal mode $s_1$, etc. The general form of state (\ref{SPDC-state}) in terms of photon momentum states can be obtained through standard SPDC analysis, see for example in Ref.~\cite{lahiri2015theory}. 

Idler photon 1 then hits the object $O$, which has a distribution of real transmittance coefficient $T(x,y)$ in the transverse plane and will transmit the idler photon in mode $i_1$ with probability $|T(x,y)|^2$ and reflect (or absorb) it with probability $|R(x,y)|^2=1-|T(x,y)|^2$. For convenience, we will omit the spatial dependence label $(x,y)$ in the following discussions and simply write the transmittance and reflection coefficients as $T$ and $R$ respectively. Then the general photon state becomes
\begin{equation}
    \label{SPDC-stateO}
    |\psi_2\ra =  c_1\ket{s_1}(T\ket{i_1}+R|r\ra)+  e^{-i\phi_2} c_2\ket{s_2}\ket{i_2} ,
\end{equation}
where $\phi_2$ is the new relative phase after $O$, and $|r\ra$ represents the mode being reflected or absorbed by $O$.

Under assumption (b), the perfect alignment of the two idler modes $\ket{i_1}$, $\ket{i_2}$, indicated by the overlapping of the two green lines after NL2, will converge into the same mode $\ket{i_0}$, which lead to the photon state to become
\beqa
\label{induce-stateO}
\ket{\psi_3} =  c_1\ket{s_1}(T\ket{i_0} + R\ket{r})+  e^{-i\phi_3} c_2  \ket{s_2}\ket{i_0},
\eeqa
where $\phi_3$ is the relative phase of the two signal photon modes before the second BS. 

Due to partial transmission of the idler mode $\ket{i_1}$ into mode $|i_0\ra$, the relative probability of the signal photon detection can be computed as 
$P = \abs{c_1 T}^2 + \abs{c_1 R}^2 + \abs{c_2}^2 + 2\abs{c_1}\abs{c_2}T \cos(\phi_3)$. Then one can obtain the interference visibility as $ V = 2\abs{c_1}\abs{c_2}\abs{T}$. Combining with the predictability $D=| \abs{c_1}^2-\abs{c_2}^2|$, one immediately achieves an imaging duality ellipse relation

 \begin{equation}
    \frac{V^2}{T^2}+D^2 =1.
    \label{IDE}
\end{equation}
Obviously, the restriction between waveness and particleness is now controlled by the object information via transmittance $T$. The partial transmission of the object serves as generalized form of partial coherence $\gamma$. Hence it connects directly to the ellipticity $\eta=1-|T|$ and the reflection coefficient $R$ is exactly the eccentricity of the ellipse. Here the existence of the object $O$ with non-perfect transmittance $T<1$ inevitably prevents perfect induction of coherence. Therefore, only partial coherence is restored from the pump photon to the signal photon. 

\begin{figure}[ht]
    \centering
    \subfigure{\includegraphics[scale=0.3]{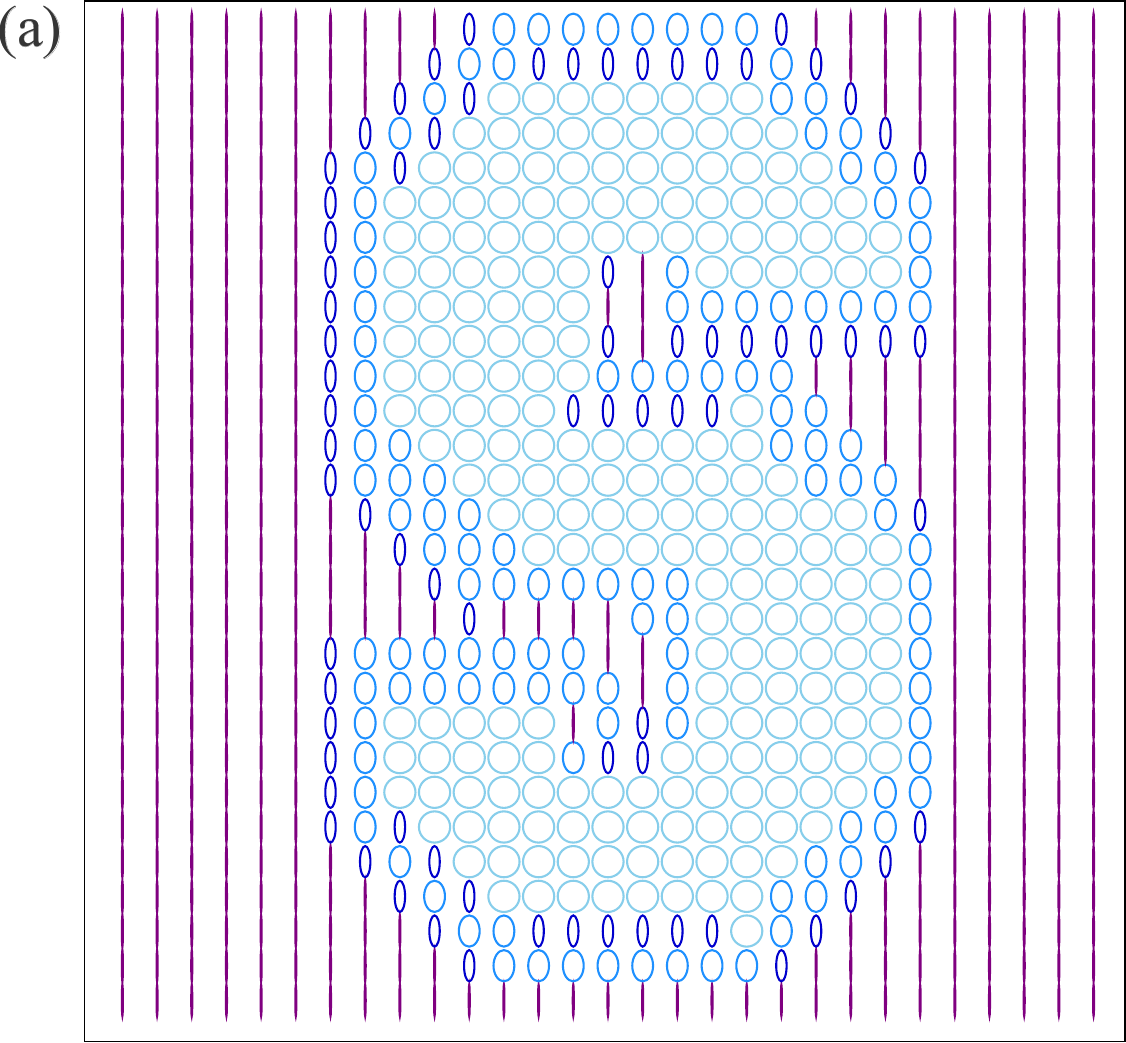}\label{fig:slogo_clean}}\qquad
    \subfigure{\includegraphics[scale=0.3]{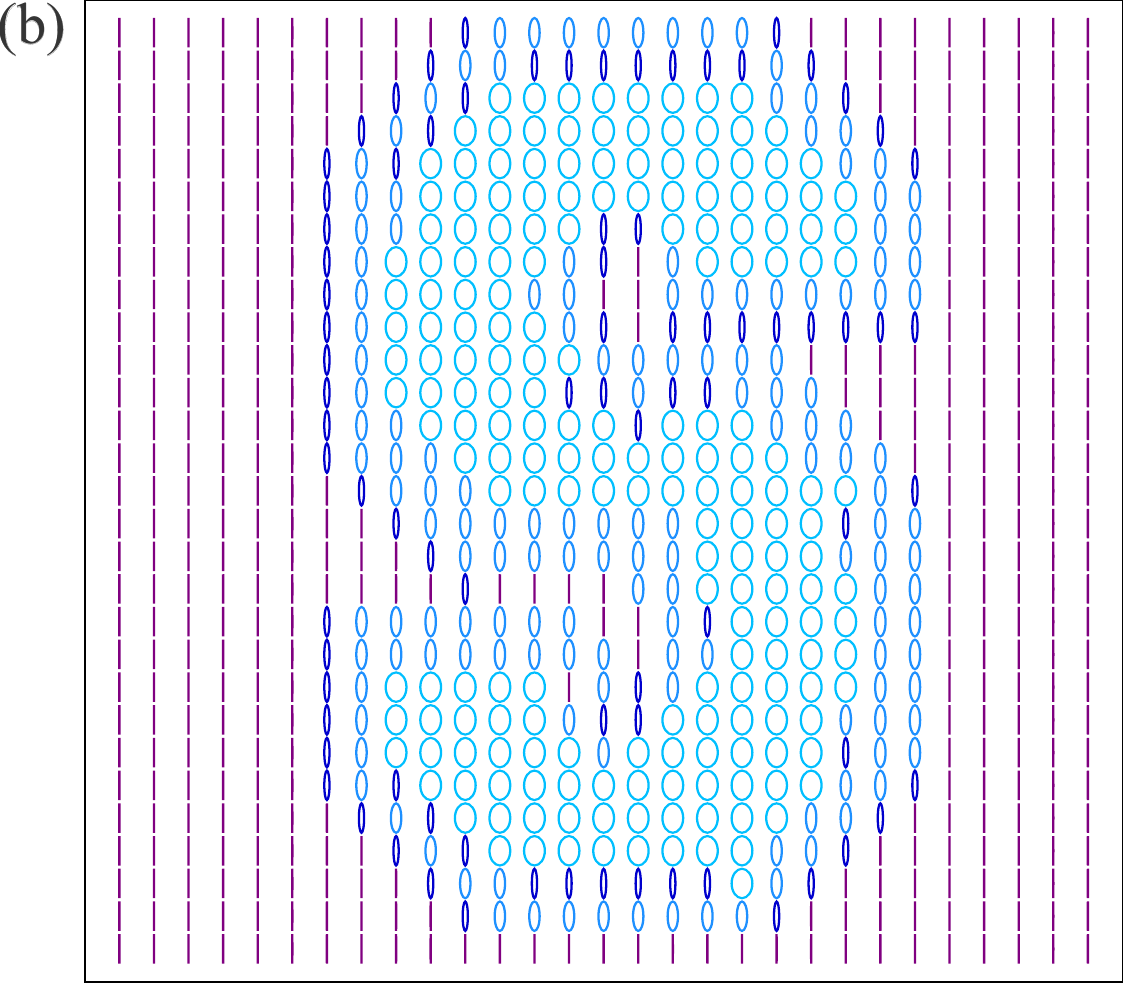}\label{fig:slogo_noisy}} 
    \caption{Illustration of the point by point duality D-V relation in the transverse plane. The blue circles represent full transmission $T=1$ cut-out parts of the object, and the blue ellipses represent the object's partial transmission $T<1$. The purple ellipses represent opaque area of the object with $T=0$. Here, the top and bottom plots represent the ideal noiseless and the non-perfect alignment practical cases respectively.}
    \label{fig:slogos}
\end{figure}

Conversely, the IDE relation above also indicates that the waveness and particleness control the image information in this QIUP scenario. {  When the visibility $V$ and predictability $D$ can be obtained by measurements, the information or image (via transmittance $T$) of the object can be determined straightforwardly.} In other words, the image of the object can be directly achieved by measuring the ellipticity of the IDE for each point in transverse plane. As an example, Fig.~\ref{fig:slogo_clean} illustrates the simulated image of an object with a cut-out shape of the letter ``S''. In this case, the transmittance $T(x,y)$ is 1 at the center of the cut-out spaces, and it gradually decreases to 0 as it gets closer to the edges of the shape. This simulated distribution of the transmittance is demonstrated by the variation of the ellipticity $\eta(x,y)=1-T(x,y)$ of the IDE in Fig.~\ref{fig:slogo_clean}. Therefore, the measurement of wave-particle dual nature $V(x,y)$ and $D(x,y)$ for each transverse spatial point is capable to retrieve the object information. It has the advantage of being robust against decoherence, non-perfect alignment, etc., and will be discussed in the following.  


{  \section{Robust Imaging Duality Ellipse Against Practical Imperfections}
In reality, there are many other sources (internal or external) of non-perfections or loss of coherence just as discussed in obtaining the general DE relation (\ref{Duality Ellipse}). Therefore, we will release the two assumptions (a), (b) adopted in the previous section and show, surprisingly, that although these non-perfections will change the IDE relation \eqref{IDE} quantitatively, they will not affect its form, thus still allowing a reliable robust way of determining the image of the object. }

By releasing assumption (a), one needs to take into account possible decoherence of two signal photon modes $|s_1\ra$ and $|s_2\ra$, or the initial non-perfect partial coherence of the pump photon modes $|1\ra$ and $|2\ra$. Both situations can be modeled effectively by introducing an external party $m$ with states $|m_1\ra$, $|m_2\ra$. As an illustration, here we discuss the case when partial coherence is originated from the pump photon without loss of any generality. That is, we consider a more general pump photon state
\beq
\ket{\psi'_0}=c_1\ket{1}|m_1\ra + c_2 e^{i\phi'_0} \ket{2}|m_2\ra,
\eeq
where $|\la m_1|m_2\ra|=\gamma <1$ is the source of partial coherence due to environmental noise or other types of decoherence mechanisms. 

By releasing assumption (b), we consider practical cases when there is misalignment of the two idler modes $i_1$, $i_2$, which will result in partial transfer of coherence from the pump photon to the signal photon \cite{zou1991induced}. That is, the idler modes after the alignment becomes, $|i_1\ra \rightarrow |i_0\ra$ but $|i_2\ra \rightarrow |i_0'\ra$, with $|\la i_0|i_0'\ra|=\alpha<1$, a non-unity overlap. 

Consequently, the down-converted photon state after idler mode alignment needs to be modified 
\beqa
  \label{practical-induce-stateO}
\ket{\psi'_2} &=& c_1\ket{s_1}(T\ket{i_0} + R\ket{r})|m_1\ra \notag\\
&&+ c_2 e^{i\phi'_2}  \ket{s_2}\ket{i'_0}|m_2\ra, 
\eeqa
where $\phi'_2$ is the generic two path phase difference. Then the probability of detecting the signal photon can be computed as $P = \abs{c_1}^2 + \abs{c_2}^2 + 2\abs{c_1}\abs{c_2}T\gamma\alpha \cos(\phi'_2)$. Straightforward calculation leads immediately to the following modified imaging duality ellipse relation
\begin{equation}
    \frac{V^2}{\gamma^2T^2\alpha^2}+D^2 =1,
    \label{generalIDE}
\end{equation}
which is again an ellipse equation with respect to $V$ and $D$. The ellipticity, however, is now modified to be $\eta=1-T \alpha \gamma$. That is, the wave-particle duality is further controlled by the two sources of partial coherence, i.e., the non-perfect idler photon alignment $\alpha$ and the original pump photon partial coherence $\gamma$. 

Although the current ellipticity can't distinguish the contributions from which of the three sources, $T$, $\alpha$, $\gamma$, {\em the above duality relation (\ref{generalIDE}) can still be used to determine the image of the object.} Apparently, the idler photon non-perfect alignment $\alpha$ and the pump photon's partial coherence $\gamma$ have a uniform effect across the entire transverse plane of the signal photon. Thus, $\alpha$ and $\gamma$ are constants for a given pump source and a systematic alignment technique, independent of the signal photon's transverse plane distribution. Consequently, these imperfections will increase the ellipticity $\eta (x,y)=1-T(x,y) \alpha \gamma$ everywhere across the transverse plane by a common factor, and $\eta (x,y)$ still has a one to one correspondence to the transmittance $\abs{T(x,y)}$. Therefore, the ellipticity $\eta (x,y)$ distribution is still an image of the object. 

Fig.~\ref{fig:slogo_noisy} illustrates the same image of the letter ``S'' as in the previous section but with the effect of the non-perfections $\alpha$, $\gamma$. The ellipticity of (\ref{generalIDE}) for all points have increased in magnitude, but the relative ellipticity differences among different spatial points remain the same. As is shown, it is still capable to map out the object information even under unknown practical noisy situations. 


It is also worth to mention that the overall unknown noise effects $\alpha \gamma$ can be obtained by measuring the ellipticity of (\ref{generalIDE}) without an object $T=1$ through the relation $\eta (x,y)=1-\alpha \gamma$. The exact transmittance distribution $T(x,y)$ can then be achieved when the object is in place.

\section{Discussion}
In conclusion we have carried out a quantitative analysis of coherence effect on wave-particle duality. It is found that an exact compact duality ellipse relation is achieved where the ellipicity is uniquely controlled by the two-path degree of coherence. This DE clarifies the ambiguity of non-exclusiveness in the previous duality inequality, and is completely compatible with Bohr's complementarity principle. 

We further extend our analysis to the application of quantum imaging with undetected photons. One is then led to a compact imaging duality ellipse equation where the object transmittance information $T$ is directly connected to ellipticity $\eta=1-T$ of the $V$-$D$ ellipse. The existence of the object amounts to an equivalent way of detecting the idler photon in mode $|i_1\ra$, obtaining the particleness information of the signal photon (compromised by reduction of wave information). Thus, achieving the object information is an equivalent way of knowing the dual nature of both wave and particle information as reflected by the imaging duality ellipse (\ref{IDE}). The compact IDE relation provides an alternative way of achieving object information through measurements of the photon's waveness and particleness in the image plane. More importantly, as shown in (\ref{generalIDE}), this technique is independent of practical non-perfections of the systems in terms of misalignment and decoherence.

{While our analysis of the duality ellipse relation \eqref{Duality Ellipse} based on photons, the results are valid for all quantum particles. Further, the DE framework can be generalized to broader quantum technologies, such as metrology, sensing, and communication, where advanced coherence measures and the interplay of wave-particle duality are central. This extension highlights the universality of coherence-modified exact complementarity, offering a unified lens to optimize protocols that leverage quantum coherence and duality.}

{Finally, we remark that the achieved duality ellipse relations \eqref{Duality Ellipse}, \eqref{generalIDE} are valid in two-path interference scenarios. It would be very interesting to explore its extension to multi-path interference, where generalized fundamental definitions of coherence, waveness and particleness \cite{durr2001quantitative, bera2015duality,basso2022entanglementpla} in the multi-path scenario will be crucial.} \\

\noindent\textbf{Acknowledgments.}
We acknowledge valuable discussions with G.S. Agarwal, and financial supports from NSF Grant No. PHY-2316878 and from Stevens Institute of Technology.


\bibliography{3references}

\end{document}